\documentclass[10pt,preprint,a4paper]{aastex}
\usepackage{graphics,epsf}
\usepackage{amsmath}                
\usepackage{amsfonts}               
\usepackage{amssymb}                
\usepackage{epsfig}                 

\def \cm{~\rm{cm}}
\def \s{~\rm{s}}
\def \km{~\rm{km}}

\def \K{~\rm{K}}

\def \AU{~\rm{AU}}
\def \erg{~\rm{erg}}

\def \yr{~\rm{yr}}

\def \kpc{~\rm{kpc}}

\begin{document}

\title{THE INTERACTION OF THE ETA CARINAE PRIMARY WIND WITH A CENTURY OLD SLOW EQUATORIAL EJECTA}

\author{Noam Soker\altaffilmark{1} and Amit Kashi\altaffilmark{2}}

\altaffiltext{1}{Department of Physics, Technion -- Israel Institute of Technology, Haifa 32000, Israel; soker@physics.technion.ac.il.}
\altaffiltext{2}{Department of Physics and Astronomy, University of Nevada, Las Vegas, 4505 S. Maryland Pkwy, Las Vegas, NV, 89154-4002, USA; kashia@physics.unlv.edu}

\begin{abstract}
We argue that the asymmetric morphology of the blue and red shifted components of the outflow
at hundreds of AU from the massive binary system $\eta$ Carinae can be understood from the
collision of the primary stellar wind with the slowly expanding dense equatorial gas.
Recent high spatial observations of some forbidden lines, e.g. [Fe~III]~$\lambda$4659, reveal the outflowing
gas within about one arcsecond ($2300 \AU$) from $\eta$ Car.
The distribution of the blue and red shifted components
are not symmetric about the center, and they are quite different from each other.
The morphologies of the blue and red shifted components correlate with the location of dense slowly moving
equatorial gas (termed the Weigelt blob environment; WBE), that is thought to have been ejected during
the 1887 -- 1895 Lesser Eruption (LE).
In our model the division to the blue and red shifted components is caused by the postshock flow of the primary wind
on the two sides of the equatorial plane after it collides with the WBE.
The fast wind from the secondary star plays no role in our model for these components, and it is the freely expanding
primary wind that collides with the WBE.
Because the line of sight is inclined to the binary axis, the two components are not symmetric.
We show that the postshock gas can also account for the observed intensity in the [Fe~III]~$\lambda$4659 line.
\end{abstract}

\keywords{stars: mass loss --- stars: winds, outflows --- stars:
variables: other --- stars: individual (Eta Car)}

\section{INTRODUCTION}
\label{sec:intro}

The colliding-wind binary system $\eta$ Car is a special binary system in our Galaxy
(e.g., \citealt{DavidsonHumphreys1997}).
It is composed of a primary Luminous Blue Variable (LBV), a very massive star ($M_1 \ga 120 \rm{M_\odot}$)
whose on the edge of instability, and a somewhat evolved massive O-star secondary ($M_2 \ga 30\rm{M_\odot}$).
The binary system has a complicated temporal behavior and a complex circumbinary nebula.
Both are connected to the massive and energetic winds of the system at present and in the past.
Interesting dramatic periodic changes, termed collectively as ``the spectroscopic event'', take place every $\sim 5.54$ years
close to periastron passage.
The effect of the secondary gravity when the secondary approaches the primary on its highly eccentric
($e \ga 0.9$) orbit is thought to be strongly related to the occasion of the spectroscopic event.
During the spectroscopic event many emission and absorption lines and bands show considerable changes for a few weeks (\citealt{Damineli2008} and references therein),
e.g., a deep minimum in the X-ray emission.
The X-ray emission, for example, comes from the colliding stellar winds,
(\citealt{Corcoran2005}; \citealt{Corcoran2010}; \citealt{Parkin2011}; \citealt{Akashi2006}, \citeyear{Akashi2011}; \citealt{Moffat2009};
\citealt{Henley2008}; \citealt{Okazaki2008}; \citealt{Pittard2002}, \citeyear{Pittard1998}; \citealt{Behar2007}; \citealt{Teodoro2012}),
while its minimum is attributed to the suppression of the secondary wind near periastron passages.
This suppression occurs when the colliding-wind region comes very close to the secondary star as the two stars approach each other,
and the secondary gravity acts on the dense postshock primary wind (\citealt{KashiSoker2009b}).
This accretion is aided by the formation of clumps in the primary wind (\citealt{Akashi2011}).
Another explanation (which requires our less favorable orientation; see below) is that the colliding-wind region is occulted by the primary (\citealt{Corcoran2010}).
However, occulation alone cannot account for the X-ray suppression (\citealt{KashiSoker2009a}; \citealt{Hamaguchi2007}).

Apart from the colliding-wind region there is a more extended structure around the binary system.
The complexity of the structure poses a challenge in understanding observations from the system,
and how different parts of it were formed.
The bipolar Homunculus nebula surrounding the system is the most prominent structural feature.
It was formed during the 1837.9 -- 1858 Great Eruption (GE) when the primary LBV ejected $12$ -- $40 \rm{M_\odot}$,
a considerable part of its mass
(\citealt{Gomez2006}, \citeyear{Gomez2010}; \citealt{SmithOwocki2006}; \citealt{Smith2003} \citealt{SmithFerland2007}; \citealt{KashiSoker2010}).
Both the GE and the weaker 1887.3 -- 1895.3 Lesser Eruption (LE) that followed it are the source of most of the circumbinary medium.
The LE was a much less energetic eruption (\citealt{Humphreys1999})
and only $0.1$ -- $1 \rm{M_\odot}$ were ejected from the primary (\citealt{Smith2005}).
Part of this material was ejected inhomogeneously in equatorial directions.
Material from the GE in the equatorial direction created the equatorial skirt, while material ejected from the LE is thought to be
the source of dense blobs and the gas in their surroundings closer to the binary system (\citealt{Smith2004}).
These blobs are known as the Weigelt blobs (WBs; \citealt{WeigeltEbersberger1986}; \citealt{HofmannWeigelt1988}).
The WBs are concentrated on the NW side of $\eta$ Car, but there is also dense material in other directions
(e.g., \citealt{Smith2000}; \citealt{Dorland2004}; \citealt{Chesneau2005}; \citealt{Gull2009}; \citealt{Mehner2010}).

In this paper we focus on the interaction of the primary wind with
material in the Weigelt blob environment (hereafter WBE).
Observations of ionized iron lines from the system in recent years provide us with detailed information about the WBE.
{}From HST/STIS spectra of Fe~II, [Fe~II], [Ni~II] and [Ni~III] lines from WBs C and D,
\cite{Smith2004} found the radial velocity of the WBs to be $\sim 40 \km \s^{-1}$.
 {}From that \cite{Smith2004} concluded that the WBs were originated in the LE.
On the other hand, \cite{Dorland2004} presented HST observations of WBs C and D together with simulations of their propagation and
concluded that they formed during the period 1910 -- 1942, preferably during a brightening event which took place in 1941.
For the purpose of this paper it is immaterial when the dense WBE was ejected.

\cite{Gull2009} presented observations of broad ($\sim 500 ~\rm{km}~\rm{s}^{-1}$) [Fe~II] emission line structures which extend to
$\sim 1600 \AU$ from the binary system, and [Fe~III], [Ar~III], [Ne~III] and [S~III] lines which extend up to a projected distance of
only $\sim 700 \AU$ from NE to SW.
The latter lines showed radial velocities with tendency to the blue ($\sim -500$ to $+200 ~\rm{km}~\rm{s}^{-1}$).
\cite{Gull2009} observed all those forbidden lines disappearing during the spectroscopic event.
\cite{Madura2012} took HST/STIS spectra of [Fe~III] emission lines from slits in different position angles through the central source
of $\eta$ Car.
The observations covered the 2003.5 spectroscopic event of $\eta$ Car.
While the low velocity component of these lines was associated with the WBs,
\cite{Madura2012} interpreted the high velocity components of these lines as being formed in the wind collision zone of the binary system.
\cite{Gull2011} presented more HST/STIS high-ionization forbidden-line observations taken after the 2009 event,
this time in the form of complete intensity maps.
The high velocity components were evident as well.
Based on comparison with SPH simulations and under the assumption (which we will dispute below) that the lines
originate in the binarycolliding-wind region, \cite{Gull2009}, \cite{Madura2011}, \cite{Madura2012} and \cite{Gull2011}, concluded that
the high velocity components of the forbidden-line observations fit an orientation in which the secondary
star is in the direction of the observer for most of the binary orbit.
Namely, the longitude angle (the argument of periapsis) is $\omega \simeq +270^\circ$ according to their interpretation.

There are several other papers claiming a longitude angle of $\omega \simeq +240^\circ$ -- $+270^\circ$
(e.g., \citealt{Moffat2009}; \citealt{Okazaki2008}; \citealt{Henley2008}; \citealt{Hamaguchi2007};
\citealt{Nielsen2007}; \citealt{Gull2011}; \citealt{Madura2010PhDT}; \citealt{Madura2011}, \citeyear{Madura2012}; \citealt{MaduraGroh2012}; \citealt{Madura2012}; \citealt{Parkin2009}).
Some other papers claim an opposite view, namely $\omega \simeq 90^\circ$
(e.g., \citealt{AbrahamGoncalves2007}; \citealt{FalcetaGoncalves2005}; \citealt{KashiSoker2008, KashiSoker2009a}).
In \cite{KashiSoker2008} we studied a variety of observations of the system,
and concluded that the longitude angle is $\omega \simeq 90^\circ$.
Among the observations that led to this conclusion was the observation of the high-excitation narrow [Ar~III]~$\lambda$7135 line,
that originates in the WBs, across the 2003.5 spectroscopic event (\citealt{Damineli2008}).
The high-ionization nature of the [Ar~III]~$\lambda$7135 line implies that the ionizing
radiation must be supplied by the hot secondary O-star, rather than by the primary star.
By comparing the rate of the companion's ionizing photons that reach WB D as a function of the orbital phase to the observed intensity of this line,
we showed that a match can only be obtained for a longitude angle of $\omega \simeq 90^\circ$.

\cite{Mehner2010} also presented HST/STIS spectra of high ionization lines ([Fe~III] and [N~III])
around $\eta$ Car extending over more than a full cycle.
They also found that most of the emission arrives from the WBs.
Moreover, they found a variation of the intensity of the lines over the entire orbital cycle,
in addition to the fast increase before periastron
that followed by the typical decrease to the spectroscopic event.
\cite{Mehner2010} identified a fast blueshifted component of the high ionization lines that appears concentrated near the
stars and elongated perpendicular to the bipolar axis of the system.
\cite{Mehner2010} suggested that this component is related to the equatorial outflow and/or to
dense material known to exist along our line of sight to the system.

Our motivation to discuss the WBE and its interaction with the primary wind is discussed in section \ref{sec:motivation}.
In section \ref{sec:Primary} we argue that the behavior of some recently observed forbidden lines matches that expected from the
primary wind interaction with the WBE.
Our discussion and summary are in section \ref{sec:summary}.

\section{MOTIVATION}
\label{sec:motivation}

Two directions motivate us to discuss the WBE and its interaction with the primary wind.
First, as the Weigelt blob environment (WBE) moves very slowly relative to the primary,
the primary wind is strongly shocked when it hits the WBE.
The WBs are located in the equatorial plane in the side closer to the observer.
In our preferred orientation of $\omega \simeq 90 ^\circ$ (\citealt{KashiSoker2008}) the primary wind
is constantly shocked when colliding with the WEB.
In the $\omega \simeq 270 ^\circ$ case, where the secondary is toward us during most of the time,
the secondary tenuous and faster wind hits the Weigelt blobs during most of the time.
However, even in that case some slower and denser primary wind gas flows toward us.
More than that, the WBE extends tangentially \citep{Chesneau2005}, over an angle larger than that covered
by the secondary wind after it collides with the primary wind.
Hence the undisturbed primary wind directly collides with part of the WBE even in models with $\omega \simeq 270 ^\circ$.
We are motivated to study what are the signatures of the wind-WBE collision and
why it is not commonly considered when modelling emission from that region.

The second direction that motivates us comes from our view that a simple model that ignores the wind-WBE collision
fails to account for the behavior of the [Fe~III]~$\lambda$4659 forbidden line from the WBE.
We refer to the model that was presented by \cite{Madura2011}, and later used by \cite{Gull2011} and \cite{Madura2011},
who considered only the collision between the two stellar winds and their emission from an extended region up to $\sim 1500 \AU$,
but not the wind-WBE collision.
The observations show forbidden line emission from a projected region of $\ga 1000 \AU$ that encloses the
WBs that reside at real distances of up to $\sim 850 \AU$ from the center (\citealt{Dorland2004}; \citealt{Chesneau2005}),
assuming the WBs are in the equatorial plane.
Both \cite{Madura2012} and \cite{Gull2011} compare their results with 3D SPH simulations of the
stellar colliding winds (see also \citealt{Gull2009}), and attribute the low velocity component
of the [Fe~III]~$\lambda$4659 line (and some other forbidden lines) to the WBs,
while the high velocity components are considered to result from the outflowing colliding stellar winds.
They then argue that a good fit is obtained only for $\omega \simeq 270^\circ$.

\begin{figure}[!t]
\resizebox{1\textwidth}{!}{\includegraphics{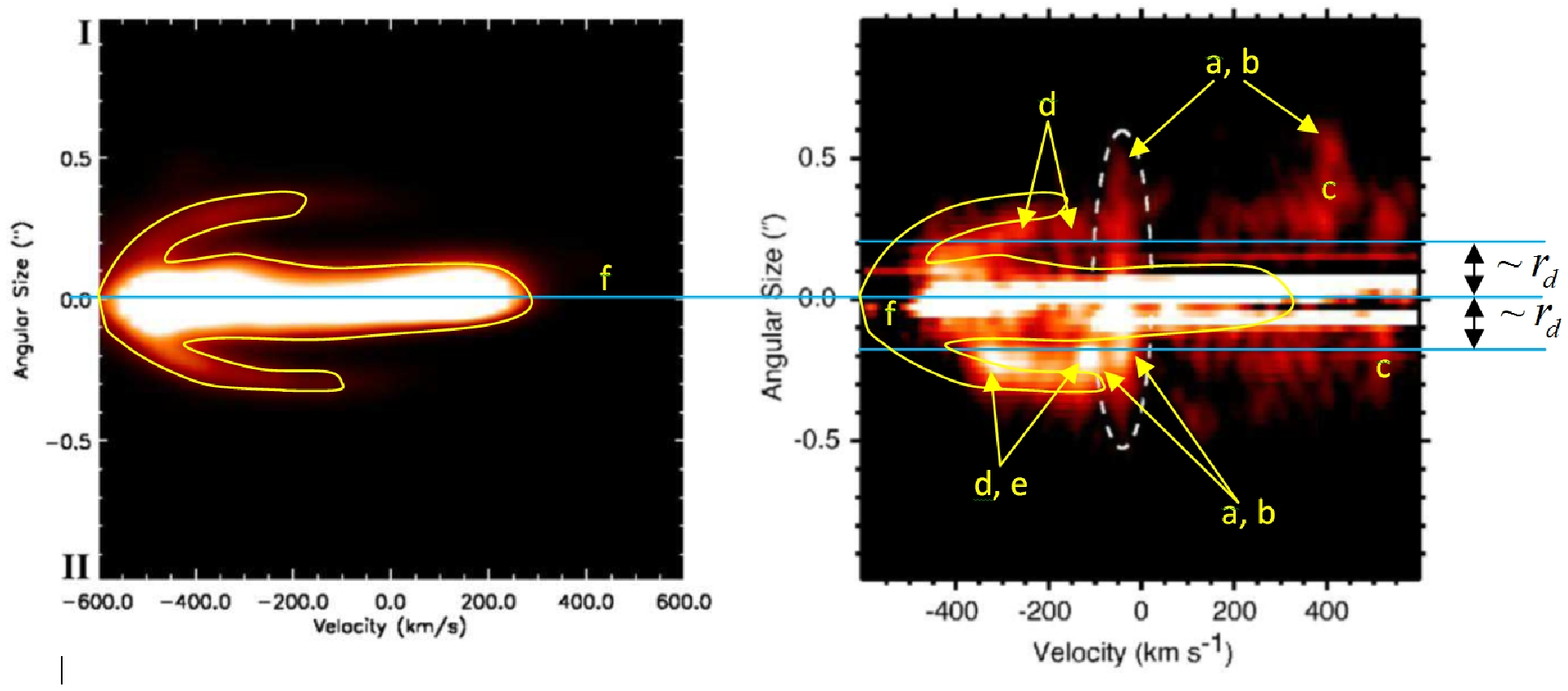}}
\caption{\footnotesize
Comparing observations (right panel) with the model (left panel) proposed by
(\citealt{Madura2011}; a similar figure is in \citealt{Madura2012}).
This figure is taken from figure 4 of \cite{Madura2011}, with our marking added.
Presented is the velocity profile along a slit more or less along the equatorial plane
at phase 0.976. The vertical axis is the position along the slit relative to the center,
where up is north-east.
The horizontal axis is the velocity. The different features are discussed in the text.
}
\label{fig:FromMadura}
\end{figure}

We find their model to fit the observations unsatisfactory.
We first examine the observation in which the spectroscopic slit is positioned more or less along the equatorial plane
as presented by \cite{Madura2011} at phase 0.976 (48 days before the 2003.5 periastron passage).
The observations are on the right panel and their best model is on the left panel of Fig. \ref{fig:FromMadura}.
In each panel the vertical axis is the position along the slit relative to the center, where up is north-east.
The horizontal axis is the velocity.
The distance $r_d$ marked on the figure is the shock distance in our proposed model.
We mark it here to emphasize that in our model the distance of the extended region from the center is determined by where the
primary wind is shocked against old ejected gas.
This is discussed in section \ref{sec:Primary}.

We note these discrepancies and puzzling properties.
The letters correspond to those we mark on Fig. \ref{fig:FromMadura}.
(a) \cite{Madura2011} attribute the zero velocity component to the Weigelt blobs, the elliptical vertical white-dashed line
marked by them in the right panel, while the fast components in their model
come from the primary wind.
It is not clear why two different components in their model, the primary wind and the Weigelt blobs, show very similar intensity.
(b) It is not clear why the two different components are at about the same distance from the center and have about the same size.
(c) There are observed redshifted components that their model does not reproduce.
\cite{Madura2012} discuss this in length (their appendix A; see also \citealt{Madura2010PhDT}).
They examine the case that the red components are contaminated.
There is definitely a large contamination of the [Fe~III]~$\lambda$4659 red part by the [Fe~II]~$\lambda$4665 line.
However, we note that part of  the red [Fe~III]~$\lambda$4659 emission is also seen in the lines of
[Fe~III]~$\lambda$4702 and [NII]~$\lambda$5756 in the images presented by \citet{Gull2011},
and in [Fe~III]~$\lambda$4702 also in \citet{Madura2012} (see also \citealt{Madura2010PhDT}).
Therefore this redshifted component seems to be real, even if much weaker than the blue shifted component and
being heavily contaminated.
(d) The observed shape of the extended emission is not as their model predicts.
The observed extended emission is at about the same distance from the center,
rather than having a larger distance toward lower velocities as their model predicts.
(e) The observed intensity does not decrease toward lower velocity as their model predicts.
(f) The observed velocity of the central part has a more extended redshifted component than a blueshifted one.
\cite{Madura2011} model has an opposite behavior.

\begin{figure}[!t]
\resizebox{1\textwidth}{!}{\includegraphics{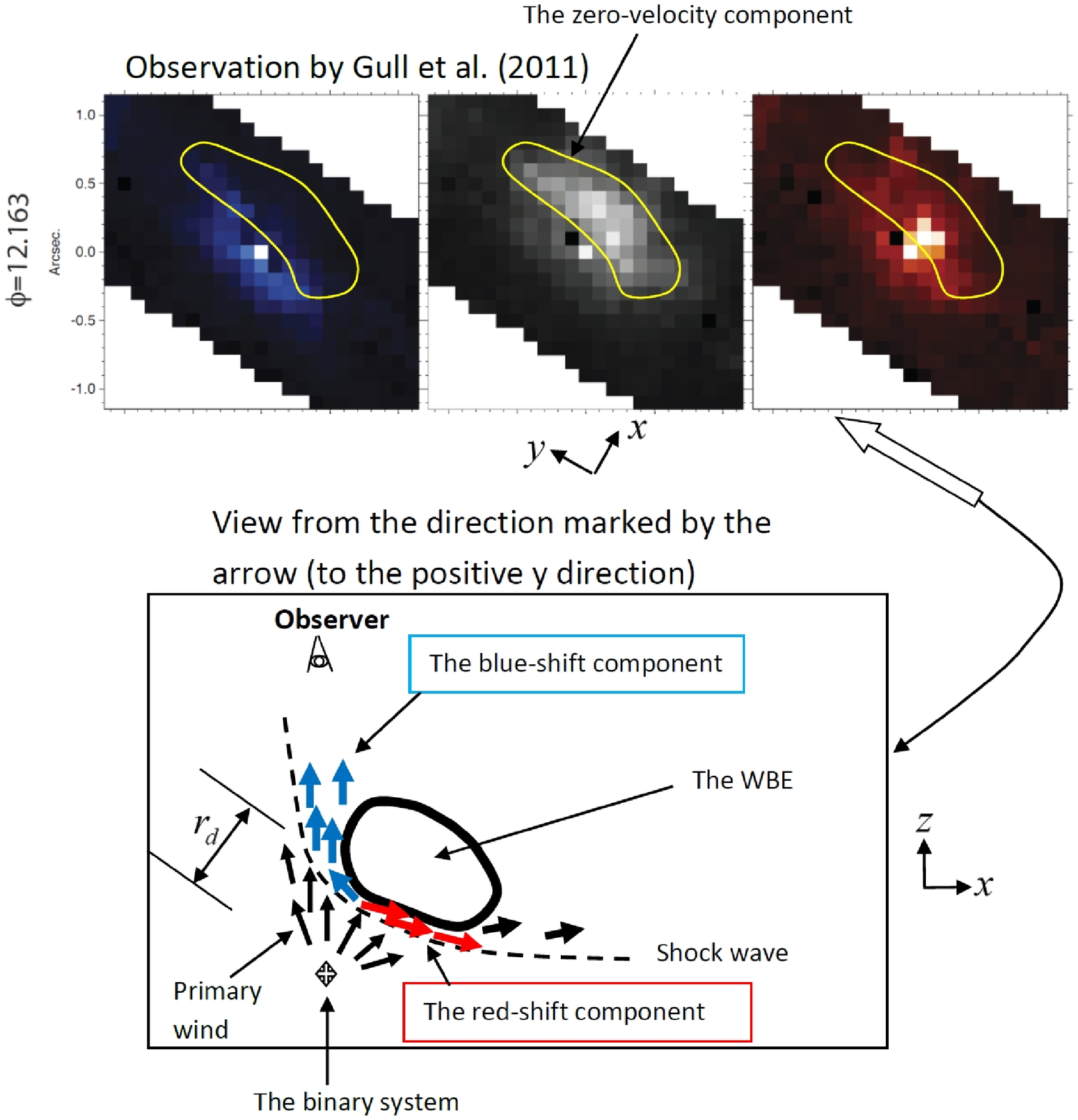}}
\caption{\footnotesize
Upper panel: The [Fe~III]~$\lambda$4659 forbidden line from \cite{Gull2011}.
The emission is presented in three velocity components:
blueshifted component ($-400$ to $-200 ~\rm{km}~\rm{s}^{-1}$), zero-velocity component ($-90$ to $-30 ~\rm{km}~\rm{s}^{-1}$)
and red-shifted component ($+100$ to $+200 ~\rm{km}~\rm{s}^{-1}$).
Bottom panel: View from the direction marked by the white arrow, according to our model.
In this model the emission comes from the shocked primary wind, that collides with the WBE ---
the dense slowly moving gas near the equatorial material.
The flow direction is schematically drawn, as well as the location of the observer according to our model.
The axes are defined in the figure:
The x-y plane is the projected view on the sky where the x-axis is more or less along the projected axis of the Homunculus on the sky.
The z-axis is directed to the observer.
}
\label{fig:FromGull}
\end{figure}

The blueshifted component of the [Fe~III]~$\lambda$4659 line is extended in the NE-SW direction and through the center.
The zero-velocity component is extended in the same direction but to the NW of the center (\citealt{Mehner2010}; \citealt{Gull2011};
see upper panel in Fig. \ref{fig:FromGull}).
The two regions seem to run along each other in the same direction and overlapping.
The redshifted component is on the NW side of the center, with its length
as the width of the zero-velocity component.
The zero velocity and blueshifted components seem to know about each other, i.e. they are correlated.

Like \cite{Madura2011}, \cite{Gull2011} and \cite{Madura2012} attribute the zero-velocity component to the WBs and the
blue and red shifted components to the colliding stellar winds.
As the zero velocity and blue shifted components seem to share some structural features, they should have the same origin,
and not come from independent components as suggested by \cite{Madura2011}, \cite{Gull2011} and \cite{Madura2012}.
It therefore seems that the interpretation of the [Fe~III]~$\lambda$4659 line (\citealt{Madura2011}, \citeyear{Madura2012} \citealt{Gull2011})
is problematic.

Motivated by the considerations above we propose that the blue and redshifted components of the [Fe~III]~$\lambda$4659 line
(and some other forbidden lines) are formed in the postshock gas of the primary wind when it collides with the WBE.
This explains the similar morphologies, sizes, intensities and other properties as we now turn to explain.
The schematic proposed flow structure is drawn in the lower panel of Fig. \ref{fig:FromGull}.

\section{THE PRIMARY WIND COLLISION WITH THE WBE}
\label{sec:Primary}

\subsection{The freely expanding wind}
\label{subsec:freely}

At electron densities of $\la 10^7 \cm^{-3}$ the emissivity ($\erg \s^{-1} \cm^{-3}$) of the
[Fe~III]~$\lambda$4659 forbidden line is proportional to the
Fe~III ions number density, $n_{III}$, times electron density, while at higher density the emissivity
becomes proportional to $n_{III}$ \citep{Garstang1978}.
There are several quantities that are difficult to calculate and hence introduce uncertainties in estimating the
power of the line.
These include the fraction of the iron atoms that are in the Fe~III ionization state and the temperature of the gas.
These quantities depend on the history of the gas, e.g., heating in a shock and subsequent radiative cooling, and the ionizing radiation.
\citet{Gull2011} attempted to calculate the [Fe~III]~$\lambda$4659 line brightness but derived values that are one and a half magnitude above
observed values.
We here limit ourselves to (a) show that our model can explain the power in this line, and
(b) understand the morphology.
We now turn to consider the central pixel of size
$0.1^{\prime \prime} \times 0.1^{\prime \prime} = 230 \AU \times 230 \AU$ (for a distance of $2.3 \kpc$ to $\eta$ Car)
in the observations presented by \citet{Gull2011}.

For an efficient emission of the [Fe~III]~$\lambda$4659 forbidden line the iron should be in large part in the
ionization state of Fe~III, which it turn requires a temperature of several $\times 10^3 \K \la T \la 10^5 \K$
(e.g, \citealt{Gnat2007}).
The wind is not at an exact equilibrium as it cools during
expansion. However, the regions of the wind that are relevant to us
are those that are ionized and heated by the secondary star. We
assume that in these regions the cooling is not much faster that the
recombination and ionization time scales of the relevant ions, and
hence can be considered for our purposes as being in
equilibrium.
In our model the basic process is that the emitting gas is at $\sim 10^4 \K$ and more or less in equilibrium.
At that temperature most of the iron is in the Fe~II ionization state (\citealt{Gnat2007}).
The secondary radiation then ionizes the Fe$^{+1}$ ions to form the desired Fe$^{+2}$ ions.

The central pixel has a size of $230 \AU$, reaching an average distance of $130 \AU$ from the center,
which is much larger than the wind interaction zone.
We therefore consider the density of the freely streaming primary wind (for the colliding zone contribution see below).
The LBV primary star blows a wind with a high mass loss rate of $\dot M_1 \simeq 3 \times 10^{-4} ~\rm{M_\odot \yr^{-1}}$ and
with a terminal velocity at the equator of $v_1 \simeq 500 \km \s^{-1}$ (e.g., \citealt{Pittard2002}).
We note that \cite{Gull2011} and \cite{Madura2012} take a higher mass loss rate of $\simeq 10^{-3} ~\rm{M_\odot \yr^{-1}}$.
However, their SPH simulations lack radiative cooling, and therefore their postshocked densities are underestimated.
We consider distances much larger than the orbital separation and neglect the motion of the primary around the center of mass.
The electron density of the freely streaming primary wind is
\begin{equation}
n_{e1} = 7 \times 10^6  
\left( \frac{\dot M_1}{3\times 10^{-4} ~\rm{M_\odot \yr^{-1}}} \right)
\left( \frac{r}{100 \AU} \right)^{-2}
\left( \frac{v_1}{500 \km \s^{-1}} \right)^{-1}
\cm^{-3}.
\label{eq:ne1f}
\end{equation}

We now scale quantities to estimate the power of the central pixel.
In planetary nebulae the power of the [Fe~III]~$\lambda$4659 line reaches values of $\la 0.025$ times the H$_\beta$ power
(e.g., \citealt{DelgadoInglada2009, SimonDiaz2011}).
We checked and found that such a ratio will be too low to explain the observed intensity of the
[Fe~III]~$\lambda$4659 line in $\eta$ Car.
This is explained by the small ratio of Fe~III ion number-density to hydrogen number density in planetary nebulae
$n_{III}/n_H \sim 3 \times 10^{-7}$ \citep{Garstang1978}.
In other environments the fraction can be much higher, e.g., \citet{Garstang1978} find this ratio to be $\sim 3 \times 10^{-6}$ in the Orion nebula.
We scale with this ratio, which implies that $\sim 10\%$ of the iron is in the Fe~III ionization state.
To reach this fraction the hard ionizing radiation of the secondary star is required.
The emissivity in the line at densities of $n_e \sim 10^7 \cm^{-3}$ according to Garstang et al. (\citeyear{Garstang1978};
we note the updated calculations of, e.g., \citealt{Keenan1992} but for the purpose of our estimates the
simple expressions of Garstang et al. are adequate) is
\begin{equation}
\epsilon_{III} \simeq 3 \times 10^{-13}                  
\beta_{III}(T)
\left( \frac{n_{III}}{3\times 10^{-6} n_H} \right)
\left( \frac{n_e}{10^7 \cm^{-3}} \right)^{\delta}
\erg \s^{-1} \cm^{-3},
\label{eq:eiii}
\end{equation}
where $\delta =2$ for lower densities and $\delta \rightarrow 1$ for higher densities.
The coefficient $\beta_{III}(T)$ has values of $3, 1, 0.06$ for temperatures of $2, 1, 0.5 \times 10^4 \K$, respectively.

Equations (\ref{eq:ne1f}) and (\ref{eq:eiii}) imply that outside the central pixel, where $\delta \simeq 2$, the
surface brightness of the line decreases with radius according to $S \propto r^{-3}$,
and the central pixel will be much brighter than its immediate surroundings.
We perform the integration over the emissivity from radius $r_{\min}$ to the average distance of the first pixel $130 \AU$,
and take $\delta =1.5$ as a gross average in that range.
Taking a distance of $2.3 \kpc$ to $\eta$ Car we find the flux from the central pixel to be
\begin{equation}
\begin{split}
I_{\rm center} \sim 10^{-11} &
\beta_{III}(T)
\left( \frac{N_{III}}{3\times 10^{-6} n_H} \right)
\left( \frac{\dot M_1}{3\times 10^{-4} ~\rm{M_\odot \yr^{-1}}} \right)^{1.5}
\\
&
\times \left( \frac{v_1}{500 \km \s^{-1}} \right)^{-1.5}
\ln \left( \frac{130 \AU}{r_{\rm min}} \right)
\erg \s^{-1} \cm^{-2}.
\end{split}
\label{eq:icenter}
\end{equation}
We note the dependence on the primary wind parameters.
For a mass loss as high as used by \citet{Gull2011}, for example, the flux will be $\sim 6$ times higher than the estimate here.

Had we considered the colliding-wind region emission would have been larger, both due to the higher density of
the shocked primary wind and due to the strong ionizing radiation that can increase the density of Fe$^{+2}$ ions.
The reason fir this is that the higher density implies also that the dependence on density (equation \ref{eq:eiii})
is linear ($\delta=1$), rather than having the steeper dependence of $\delta=2$.
Though the density of the shocked primary wind is higher, the volume occupied by it is smaller than if it was freely streaming.
Therefore, we expect the addition from the shocked primary wind to be only slightly more had it been
freely streaming.

The observed flux in the central pixel observed by \citet{Gull2011} is $\sim 3 \times 10^{-11} \erg \s^{-1} \cm^{-2}$
(when summing up the three velocity components).
Despite the crude derivation we can conclude that the primary wind can account for the brightness
of the central pixel as observed by \citet{Gull2011}.
The primary wind explanation accounts also for the central pixel being much brighter than
its immediate environment (not including the extended bright regions).
The model presented by \citet{Madura2012} and \citet{Gull2011} does not reproduce a bright central region that is detached from
the extended emission.

\subsection{The Weigelt Blob Environment (WBE)}
\label{subsec:Weigelt}

Consider WB D as an example.
By inspecting the relevant figures in \cite{Dorland2004}, \cite{Chesneau2005} and \cite{Gull2009},
it can be seen that WB D covers an opening angle of $\alpha_W \simeq 30^\circ$ of the primary wind.
The radius of WB D is $r_b \sim 220 \AU$, and its distance from the center is taken to be $\sim 850 \AU$.
The electron density in WB D is comparable to the density in the other blobs and estimated to be
$n_W \simeq 10^5 - 10^{10} \cm^{-3}$ (\citealt{Verner2002}).
The mass of WB D is $M_W \simeq 7 \times 10^{-4} (n_W/10^7 \cm^{-3}) ~\rm{M_\odot}$.
Later we note that for our model the much extended, somewhat lower-density region around
WB D plays the main role in colliding with the primary wind.

{}From HST/STIS spectra of Fe~II, [Fe~II], [Ni~II] and [Ni~III] lines from WBs C and D, \cite{Smith2004}
measured the  radial  velocities of these blobs, and found them to be $v_W \sim 40 \km \s^{-1}$.
With an opening angle of $\alpha_W=30^\circ$ the blob intersects a fraction of $f_w=0.017$ of the primary wind.
Over the 120 years since the Lesser Eruption this amounts to a mass of
$\Delta M_W \simeq 6 \times 10^{-4} (f_w/0.017) ~\rm{M_\odot}$
from the primary wind that collided with the blob.
\cite{Raga1998} have shown that interaction of a wind with a static molecular
clump can cause the clump material to accelerate without
being dissociated, thus producing high-velocity molecular emission.
It is therefore safe to assume that the interaction of the primary wind with the blobs can accelerate them.
The primary wind would accelerate the blob to a speed of $v_a \simeq v_1 (\Delta M_W/M_W)$.
The mass of the blob should therefore be $M_W \ga 0.01 ~\rm{M_\odot}$ in order for it not to be accelerated
to $v_a >40 \km \s^{-1}$.
The WBs are part of a more massive gas complex (\citealt{Chesneau2005}), which we term the Weigelt blob environment (WBE),
that withstands the acceleration by the primary wind.

As the WBE was not accelerated to a high velocity, we can put a lower constraint on its mean density
\begin{equation}
n_{\rm WBE,min} \sim 3 \times 10^8
\left(\frac{v_a/v_W}{10}\right)
\left(\frac{\dot M_1}{3 \times 10^{-4} ~\rm{M_\odot \yr^{-1}}}\right)
\left( \frac{r_b}{200 \AU} \right)^{-1} 
\cm^{-3}.
\label{eq:nWBEmin}
\end{equation}
We estimate the density of the central part of the WBE to be $n_{\rm WBE} \sim 10^9 \cm^{-3}$.
In the outskirts of the WBE the density will be lower than $n_{\rm WBE}$, due to interaction with the primary
wind and evaporation.
The inner edge of the WBE is closer to the binary system.
{}From images given by \cite{Chesneau2005} we estimate the average value of the distance between the inner edge of the WBE
and the binary system to be about half the distance of the center of the
WBs or less, $r_d \lesssim 400 \AU$.

\subsection{Interaction with the WBE}
\label{subsec:interaction}

To check whether the interaction between the primary wind and the WBE can produce the observed flux from the [Fe~III]~$\lambda$4659 line,
a simple estimate can be done as follows.
The inner edge of the WBE is at a distance of $r_d$ from the binary system.
The immediate post shock electron density of the primary wind is $\sim 4$ times the pre-shock electron density, and
it is given by
\begin{equation}
n_{ep1} \simeq 2 \times 10^6
\left( \frac{\dot M_1}{3\times 10^{-4} ~\rm{M_\odot \yr^{-1}}} \right)
\left( \frac{r_d}{400 \AU} \right)^{-2}
\left( \frac{v_1}{500 \km \s^{-1}} \right)^{-1}
\cm^{-3}. 
\label{eq:ne1p}
\end{equation}
The shocked gas is heated to a temperature of up to $T \simeq 3 \times 10^6$.
It cools radiatively to $\sim 10^4 \K$ within a time scale of $\sim 6$~months.
However, when it starts to cool it is compressed, and the cooling time shortens.
Overall, the gas cools over a time scale of a few months.
The flow time is $t_f > r_d/v_1 \simeq 4 \yr$, such that the post-shock gas cools quite rapidly.
The gas cools to $\sim 10^4 \K$ as it flows.
This is the time when it efficiently emits the [Fe~III]~$\lambda$4659 forbidden line
if it is ionized by the UV radiation from the companion star. A cooling by a factor of $\sim 100$ can
result in a compression by a similar factor near the stagnation point of the interaction.
However, the gas flows outward from the interaction region and the density decreases.
We scale the density with an increase by about one order of magnitude from the immediate post shock region
to $n_{ef} \sim 10^7 \cm^{-3}$.

The size of an arcsec$^2$ at a distance of $2.3 \kpc$ to $\eta$ Car is $(2300 \AU)^2$.
We take the length of the region to be $\sim r_d$.
By using equation (\ref{eq:eiii}) we find the surface brightness of the postshock primary wind
in the [Fe~III]~$\lambda$4659 line to be
\begin{equation}
S_{\rm WBE} \simeq 3 \times 10^{-9}
\beta_{III}(T)
\left( \frac{n_{III}}{3\times 10^{-6} n_H} \right)
\left( \frac{n_{ef}}{10^7 \cm^{-3}} \right)^{\delta}
\left( \frac{r_d}{400 \AU} \right)
\erg \cm^{-2} \s^{-1} {\rm arcsec}^{-2}.
\label{eq:iwbe}
\end{equation}

The maximum flux in the [Fe~III]~$\lambda$4659 line from a pixel of $0.1^{\prime \prime} \times 0.1^{\prime \prime}$
as presented by \cite{Gull2011} is $\sim 10^{-11} \erg \cm^{-2} \s^{-1}$.
\citet{Mehner2011} find the maximum power to be from WB C, with a flux of $2 \times 10^{-12} \erg \cm^{-2} \s^{-1}$
within $0.1^{\prime \prime}$ from the blob.
Our simple estimate shows that the postshock primary wind can account for the intensity of the
[Fe~III]~$\lambda$4659 forbidden line in the extended region within one arcsec from the center.
We emphasize that the postshock region in our model for the extended emission refers to the shock
formed when the primary wind collides with the WBE and not with the secondary wind.

\subsection{Geometry}
\label{subsec:geometry}

\begin{figure}[!t]
\resizebox{1\textwidth}{!}{\includegraphics{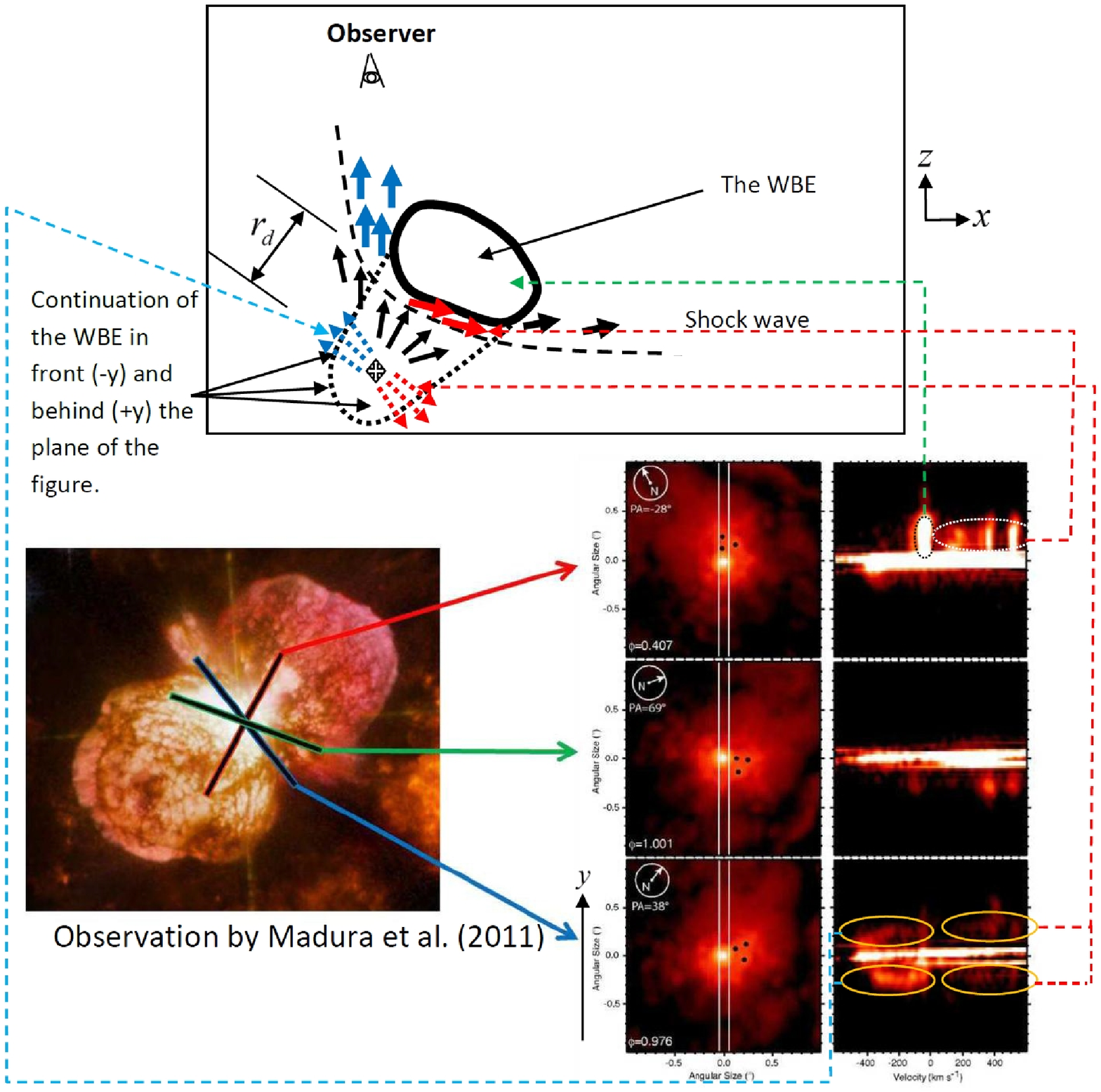}}
\caption{\footnotesize
The geometry of the WBE and the source of the [Fe~III]~$\lambda$4659 forbidden line emission.
The bottom panel, taken from \cite{Madura2011}, shows observations of the line in 3 position angles.
WBs B, C, and D are marked as black points.
The upper panel and the arrows connecting the bottom panel to the upper one show from where each of the
three components is emitted according to our model.
The zero-velocity component comes from the edge of the WBE region which is impinged by the primary wind
and ionized by the secondary ionizing radiation.
The blueshifted and redshifted components come from the postshocked primary wind gas after it cools to
$\sim 10^4 \K$ and ionized by the secondary star.
The redshifted component is contaminated \citep{Madura2012},
However, as discussed in the text, we find that a weak redshifted component does exist on top of the contamination.
}
\label{fig:Geometry}
\end{figure}

We consider the 3D geometry of the primary wind interaction with the WBE, and take into account the inclination angle and the orbital orientation.
The WBs extend towards us and to the NW side (\citealt{Dorland2004}; \citealt{Chesneau2005}; \citealt{Gull2009}).
In Fig. \ref{fig:FromGull} and \ref{fig:Geometry} we present the geometry of the WBE.
We use an axes system where the x-y plane is the plane of the sky, where the x-axis is more or less along the
projected axis of the Homunculus on the sky, and the z-axis is directed to the observer.
The observer in our model, as in our previous papers (e.g., \citealt{KashiSoker2008}, \citeyear{KashiSoker2009a})
is located at a longitude angle of $\omega=90^\circ$, namely, the primary is towards the observer for most of the orbit.
The secondary is towards the observer only for a short time close to periastron passage.
As the WBE is located between the observer and the binary system, the primary is also towards the WBE for most of the orbit.
Therefore, the WBE is exposed to the primary wind which collides with it, goes through a shock wave, and flows
around the WBE.

In our model the emission of the [Fe~III]~$\lambda$4659 forbidden line, as well as some other lines such as
the [Fe~III]~$\lambda$4702 and the [NII]~$\lambda$5756 lines, come from the shocked primary
wind, after it collides with the WBE.
One important note is in place here: the reader should be aware that usually the term `shocked primary wind'
refers to the primary wind after it is shocked when colliding with the secondary wind.
But here we refer to the shocked primary wind after it collides with the WBE.
The flow direction of the wind after this collision is schematically drawn in Fig. \ref{fig:FromGull} and \ref{fig:Geometry}.
\cite{Gull2011} divide the [Fe~III]~$\lambda$4659 forbidden line emission into three components:
blueshifted component ($-400$ to $-200 ~\rm{km}~\rm{s}^{-1}$), zero-velocity component ($-90$ to $+30 ~\rm{km}~\rm{s}^{-1}$)
and red-shifted component ($+100$ to $+200 ~\rm{km}~\rm{s}^{-1}$).
\cite{Gull2011} attribute the zero velocity component to WBs.
We accept this in our model, and more specifically we
attribute the zero velocity component to the edge of the WBE region which is impinged by the primary wind and ionized by the secondary ionizing radiation.
The WBs themselves are too dense to emit the [Fe~III]~$\lambda$4659 forbidden line, but the density at the WBE edges is low enough to allow it.
We can see this from the constraint derived above (Section \ref{subsec:Weigelt})
on the density of the WBs which is $ \sim 10^9 \cm^{-3}$, where the emission of the line is not efficient.

The blueshifted and redshifted components come from the postshocked primary wind gas after it cools to
$\sim 10^4 \K$ and is ionized by the secondary star.
The division into the blue and red shifted components is caused by the postshock flow on the two sides of the equatorial plane of
the binary system, where the WBE resides.
Because the line of sight is inclined to the binary axis the two components are not symmetric in the position-velocity diagram.
As shown in equation (\ref{eq:iwbe}) and the related discussion, this region can quantitatively explain the flux
in the emission line.

The peak radial velocity component of the forbidden emission lines is at $\sim 400 ~\rm{km}~\rm{s}^{-1}$ \citep{Gull2011},
lower than the terminal velocity of the primary wind of $\sim 500 ~\rm{km}~\rm{s}^{-1}$.
In our model the slowing down is a consequence of the interaction of the primary wind with the slowly moving WBE.
In a model where the winds are attributed to the primary wind collision with the faster secondary wind, higher velocity
than the primary wind velocity might be obtained.
Indeed, the model of \citet{Gull2011} predict blueshifted
velocities in the extended components that are $\sim 20 \%$ higher than observed (see figure \ref{fig:FromMadura}).

\section{DISCUSSION AND SUMMARY}
\label{sec:summary}

As the primary stellar wind collides with the slow dense material in the Weigelt blob environment (WBE) it is shocked.
The postshocked gas reaches a temperature of $\sim 10^6 \K$ but rapidly cools to $\sim 10^4 \K$ and is compressed.
This warm dense gas is expected to become a source of some emission lines, including forbidden lines.
The interaction of the primary wind with the WBE cannot be ignored when modelling emission from that region.
The high resolution observations reported recently by \cite{Gull2011} and \cite{Madura2012} that extend to $\sim 0.^{\prime \prime}5$
from the binary system overlap with the WBE as mapped by \citet{Chesneau2005}.
We showed that the intensity and location of the different velocity components of the [Fe~III]~$\lambda$4659
extended forbidden line \citep{Madura2011, Gull2011}, can be explained by the interaction of the primary wind with the WBE.
The geometry of our proposed model is depicted in figures \ref{fig:FromGull} and \ref{fig:Geometry}, while the
intensity is calculated in section \ref{subsec:interaction}.

In our proposed model the freely expanding primary wind continuously collides with the WBE and is shocked.
As the WBs are closer to the observer, so should be the primary most of the time.
This means that the orientation is such that $\omega \simeq 90^\circ$: the primary is closer to us
for most of the binary orbit.
The secondary becomes closer to us only for a short time near periastron passage.
This model overcomes some difficulties in the model of \cite{Madura2011} and \cite{Gull2011}
that assumes an opposite longitude angle of $\omega \simeq 270^\circ$ where it is the secondary that faces
the WBE for most of the time (see section \ref{sec:motivation}).

Our explanation of the recent observations of \cite{Madura2011} and \cite{Gull2011} with a model
where the primary star is closer to us for most of the orbit adds to a growing number of observations
that support this orientation.
The observations that support an orientation of $\omega \simeq 90^\circ$ can be summarized as follows.
\newline
(1) The Doppler periodicity of the P Cygni component of the He~I optical lines is well explained if
the origin of the absorption is taken to be in the acceleration zone of the secondary wind \citep{KashiSoker2008},
and the binary system is oriented such that $\omega \simeq 90^\circ$.
I.e., the secondary wind absorbs from the emission of the secondary photosphere.
\newline
(2) The Doppler variation of the N~II~$\lambda\lambda$5668--5712 lines observed by \cite{Mehner2011} is also well explained if they come from
the acceleration zone of the secondary wind (\citealt{KashiSoker2011}).
\cite{Mehner2011} criticized this claim by the fact that the very small locale near
the secondary star cannot produce substantial absorption lines in the spectrum of the much more luminous primary.
Our answer to that criticism was already discussed in \cite{KashiSoker2008} in the context of the He I lines.
The argument goes as follows.
The absorption in both the N~II and the He~I lines is $\la 15 \%$, and in some cases is from the excess
emission of the P Cygni lines (see observations in \citealt{Hillier2001}).
The secondary contributes $\sim 20 \%$ of the luminosity in the optical wavelengths and therefore there is no
problem for the acceleration zone of the secondary to account for the absorption intensity of these lines.
\newline
(3) The Doppler shift of the low ionization Fe~II~$\lambda$6455 line is well explained by a $\omega \simeq 90^\circ$ orientation
when it is attributed to the primary wind \citep{KashiSoker2008}.
\newline
(4) The periodic Doppler shift variation of the hydrogen Paschen lines is also explained by the same orientation
if the lines are attributed to the apex region (stagnation point) of the colliding winds \citep{KashiSoker2008}.
\newline
(5) The observed variation of the He~I~$\lambda$10830 line (\citealt{Groh2010}), in which the fast blue absorption component
exists for only several weeks prior to the periastron passage was explained in \cite{Kashi2011} using a 3D numerical simulation.
This simulation showed that the fast variation is well reproduced for $\omega \simeq 90^\circ$.
This model took into account the fact that the line is emitted from an extended source around the primary,
an ingredient that was missing in the model of \cite{Groh2010} who deduced an opposite orientation.
\newline
(6) The hydrogen column density deduced from X-ray absorption in XMM-Newton observations (\citealt{Hamaguchi2007})
could only be explained quantitatively by this orientation (\citealt{KashiSoker2009a}).
It was quantitatively shown that an orientations of $\omega = 0^\circ$, $180^\circ$ and $270^\circ$ give results
contradicting observations.
It is important to note that in contrast to the hydrogen column density, the X-ray light
curve by itself cannot be used to deduce the periastron longitude.
Minor adjustments of some parameters of the binary system (inclination, eccentricity, etc)
allow one to fit the X-ray light curve with almost any orientation (see further discussion in \citealt{KashiSoker2009a}).
\newline
(8) The variation in the intensity of the highly excited [Ar~III] narrow lines, which also originated in the WBs (\citealt{Damineli2008}),
follows the ionizing radiation from the companion for an orientation where the secondary is towards the WBs only for a short time
at periastron passage.
In section 4 of \cite{KashiSoker2008} we showed that even when behind the primary the secondary can ionize the WBE.
The reason is that the primary wind absorbs the high-ionizing secondary
radiation only within a small angle close to the primary (\citealt{KashiSoker2007}), and the WBE is extended.
The fact that in our model the primary wind absorbs a non-negligible (but not all) of the secondary ionizing radiation,
allows us to explain the variation within a cycle, and between cycles \citep{KashiSoker2008}.
On the contrary, in the model preferred by \cite{Gull2011} and \cite{Madura2012}, the secondary faces the Weigelt blobs.
This implies that the Weigelt blobs received a constant ionizing emission for most of the orbital period,
hence a constant [Ar~III] emission is expected. This contradicts observations.

The determination of the orbital orientation, i.e., whether the primary or the secondary is closer to us at periastron passage,
has implications that go beyond the goal of explaining the periodic behavior of $\eta$ Car.
Knowledge on the orientation will allow us to better understand the interaction of the binary system near periastron
passages.
The emission and absorption behavior of $\eta$ Car near periastron passage cannot be explained if the secondary wind
is not substantially suppressed then.
This suppression of the secondary wind is thought to be caused by accretion of primary wind material onto the secondary
star near periastron passage (e.g., \citealt{KashiSoker2009b}; \citealt{Akashi2011}).
Numerical simulations show indeed that near periastron passage instabilities in the wind colliding region lead to
the formation of dense blobs that are accreted toward the secondary star (\citealt{Akashi2011}; but see \citealt{Parkin2011}).
Although the requirement for the accretion process does not depend on the orientations, knowing the orientation will
help to understand the onset of, and the exit from, the several week-long accretion phase.
Understanding the accretion process in present-day $\eta$ Car might shed light on the major accretion process that took
place during the 1837-1856 Great Eruption, with implications to other LBV binary systems.

We thank one of the two referees (A. Moffat) for very helpful comments that substantially improved the manuscript.
This research was supported by the Asher Fund for Space Research at the Technion, and the Israel Science Foundation.


\newpage
\footnotesize

\begin{thebibliography}

\bibitem[Abraham \& Falceta-Gon{\c c}alves(2007)]{AbrahamGoncalves2007} Abraham, Z., \& Falceta-Gon{\c c}alves, D.\ 2007, \mnras, 378, 309

\bibitem[Akashi et al.(2011)]{Akashi2011} Akashi, M., Kashi, A., \& Soker, N.\ 2011 (arXiv:1106.2438)

\bibitem[Akashi et al.(2006)]{Akashi2006} Akashi, M., Soker, N., \& Behar, E.\ 2006, \apj, 644, 451

\bibitem[Behar et al.(2007)]{Behar2007} Behar, E., Nordon, R., \& Soker, N.\ 2007, \apjl, 666, L97

\bibitem[Chesneau et al.(2005)]{Chesneau2005} Chesneau, O., Min, M., Herbst, T., et al.\ 2005, \aap, 435, 1043

\bibitem[Corcoran(2005)]{Corcoran2005} Corcoran, M.~F.\ 2005, \aj, 129, 2018

\bibitem[Corcoran et al.(2010)]{Corcoran2010} Corcoran, M.~F., Hamaguchi, K., Pittard, J.~M., et al.\ 2010, \apj, 725, 1528

\bibitem[Damineli et al.(2008)]{Damineli2008} Damineli, A., Hillier, D.~J., Corcoran, M.~F., et al.\ 2008, \mnras, 386, 2330

\bibitem[Davidson \& Humphreys(1997)]{DavidsonHumphreys1997} Davidson, K., \& Humphreys, R. M. 1997, ARA\&A, 35, 1

\bibitem[Delgado Inglada et al.(2009)]{DelgadoInglada2009} Delgado Inglada, G., Rodr{\'{\i}}guez, M., Mampaso, A., \& Viironen, K.\ 2009, \apj, 694, 1335

\bibitem[Dorland et al.(2004)]{Dorland2004} Dorland, B.~N., Currie, D.~G., \& Hajian, A.~R.\ 2004, \aj, 127, 1052

\bibitem[Falceta-Gon{\c c}alves et al.(2005)]{FalcetaGoncalves2005} Falceta-Gon{\c c}alves, D., Jatenco-Pereira, V., \& Abraham, Z.\ 2005, \mnras, 357, 895

\bibitem[Garstang et al.(1978)]{Garstang1978} Garstang, R.~H., Robb, W.~D., \& Rountree, S.~P.\ 1978, \apj, 222, 384

\bibitem[Gnat \& Sternberg(2007)]{Gnat2007} Gnat, O., \& Sternberg, A.\ 2007, \apjs, 168, 213

\bibitem[Gomez et al.(2010)]{Gomez2010} Gomez, H.~L., Vlahakis, C., Stretch, C.~M., et al.\ 2010, \mnras, 401, L48

\bibitem[Gomez et al.(2006)]{Gomez2006} Gomez, H.~L., Dunne, L., Eales, S.~A., \& Edmunds, M.~G.\ 2006, \mnras, 372, 1133

\bibitem[Groh et al.(2010)]{Groh2010} Groh, J.~H., Nielsen, K.~E., Damineli, A., et al.\ 2010, \aap, 517, A9

\bibitem[Gull et al.(2009)]{Gull2009} Gull, T.~R., Nielsen, K.~E., Corcoran, M.~F., et al.\ 2009, \mnras, 396, 1308

\bibitem[Gull et al.(2011)]{Gull2011} Gull, T.~R., Madura, T.~I., Groh, J.~H., \& Corcoran, M.~F.\ 2011, \apjl, 743, L3

\bibitem[Hamaguchi et al.(2007)]{Hamaguchi2007} Hamaguchi, K., Corcoran, M.~F., Gull, T., et al.\ 2007, \apj, 663, 522

\bibitem[Henley et al.(2008)]{Henley2008} Henley, D.~B., Corcoran, M.~F., Pittard, J.~M., et al.\ 2008, \apj, 680, 705

\bibitem[Hillier et al.(2001)]{Hillier2001} Hillier, D.~J., Davidson, K., Ishibashi, K., \& Gull, T.\ 2001, \apj, 553, 837

\bibitem[Hofmann \& Weigelt(1988)]{HofmannWeigelt1988} Hofmann, K.-H., \& Weigelt, G.\ 1988, \aap, 203, L21

\bibitem[Humphreys et al.(1999)]{Humphreys1999} Humphreys, R.~M., Davidson, K., \& Smith, N.\ 1999, \pasp, 111, 1124

\bibitem[Kashi \& Soker(2007)]{KashiSoker2007} Kashi, A. \& Soker, N.\ 2007, MNRAS, 378, 1609

\bibitem[Kashi \& Soker(2008)]{KashiSoker2008} Kashi, A., \& Soker, N.\ 2008, \mnras, 390, 1751

\bibitem[Kashi \& Soker(2009a)]{KashiSoker2009a} Kashi, A., \& Soker, N.\ 2009, \mnras, 397, 1426

\bibitem[Kashi \& Soker(2009b)]{KashiSoker2009b} Kashi, A., \& Soker, N.\ 2009, \na, 14, 11

\bibitem[Kashi \& Soker(2010)]{KashiSoker2010} Kashi, A., \& Soker, N.\ 2010, \apj, 723, 602

\bibitem[Kashi \& Soker(2011)]{KashiSoker2011} Kashi, A., \& Soker, N.\ 2011 (arXiv:1104.4655)

\bibitem[Kashi et al.(2011)]{Kashi2011} Kashi, A., Soker, N., \& Akashi, M.\ 2011, \mnras, 413, 2658

\bibitem[Keenan et al.(1992)]{Keenan1992} Keenan, F.~P., Berrington, K.~A., Burke, P.~G., et al.\ 1992, \apj, 384, 385

\bibitem[Madura(2010)]{Madura2010PhDT} Madura, T.~I.\ 2010, Ph.D.~Thesis,

\bibitem[Madura et al.(2011)]{Madura2011} Madura, T.~I., Gull, T.~R., Owocki, S.~P., Okazaki, A.~T., \& Russell, C.~M.~P.\ 2011, Bulletin de la Societe Royale des Sciences de Liege, 80, 694

\bibitem[Madura \& Groh(2012)]{MaduraGroh2012} Madura, T.~I., \& Groh, J.~H.\ 2012, \apjl, 746, L18

\bibitem[Madura et al.(2012)]{Madura2012} Madura, T.~I., Gull, T.~R., Owocki, S.~P., et al.\ 2012, \mnras, 420, 2064

\bibitem[Mehner et al.(2010)]{Mehner2010} Mehner, A., Davidson, K., Ferland, G.~J., \& Humphreys, R.~M.\ 2010, \apj, 710, 729

\bibitem[Mehner et al.(2011)]{Mehner2011} Mehner, A., Davidson, K., \& Ferland, G.~J.\ 2011, \apj, 737, 70

\bibitem[Moffat \& Corcoran(2009)]{Moffat2009} Moffat, A.~F.~J., \& Corcoran, M.~F.\ 2009, \apj, 707, 693

\bibitem[Nielsen et al.(2007)]{Nielsen2007} Nielsen, K.~E., Corcoran, M.~F., Gull, T.~R., et al.\ 2007, \apj, 660, 669

\bibitem[Okazaki et al.(2008)]{Okazaki2008} Okazaki, A.~T., Owocki, S.~P., Russell, C.~M.~P., \& Corcoran, M.~F.\ 2008, \mnras, 388, L39

\bibitem[Parkin et al.(2009)]{Parkin2009} Parkin, E.~R., Pittard, J.~M., Corcoran, M.~F., Hamaguchi, K., \& Stevens, I.~R.\ 2009, \mnras, 394, 1758

\bibitem[Parkin et al.(2011)]{Parkin2011} Parkin, E.~R., Pittard, J.~M., Corcoran, M.~F., \& Hamaguchi, K.\ 2011, \apj, 726, 105

\bibitem[Pittard et al.(1998)]{Pittard1998} Pittard, J.~M., Stevens, I.~R., Corcoran, M.~F., \& Ishibashi, K.\ 1998, \mnras, 299, L5

\bibitem[Pittard \& Corcoran(2002)]{Pittard2002} Pittard, J.~M., \& Corcoran, M.~F.\ 2002, \aap, 383, 636

\bibitem[Raga et al.(1998)]{Raga1998} Raga, A.~C., Canto, J., Curiel, S., \& Taylor, S.\ 1998, \mnras, 295, 738

\bibitem[Sim{\'o}n-D{\'{\i}}az \& Stasi{\'n}ska(2011)]{SimonDiaz2011} Sim{\'o}n-D{\'{\i}}az, S., \& Stasi{\'n}ska, G.\ 2011, \aap, 526, A48

\bibitem[Smith(2005)]{Smith2005} Smith, N.\ 2005, \mnras, 357, 1330

\bibitem[Smith \& Gehrz(2000)]{Smith2000} Smith, N., \& Gehrz, R.~D.\ 2000, \apjl, 529, L99

\bibitem[Smith et al.(2003)]{Smith2003} Smith, N., Gehrz, R.~D., Hinz, P.~M., et al.\ 2003, \aj, 125, 1458

\bibitem[Smith\& Ferland(2007)]{SmithFerland2007} Smith, N., \& Ferland, G.~J.\ 2007, \apj, 655, 911

\bibitem[Smith et al.(2004)]{Smith2004} Smith, N., Morse, J.~A., Gull, T.~R., et al.\ 2004, \apj, 605, 405

\bibitem[Smith \& Owocki(2006)]{SmithOwocki2006} Smith, N., \& Owocki, S.~P.\ 2006, \apjl, 645, L45

\bibitem[Teodoro et al.(2012)]{Teodoro2012} Teodoro, M., Damineli,A., Arias, J.~I., et al.\ 2012, \apj, 746, 73

\bibitem[Verner et al.(2005)]{Verner2005} Verner, E., Bruhweiler, F., \& Gull, T.\ 2005, \apj, 624, 973

\bibitem[Verner et al.(2002)]{Verner2002} Verner, E.~M., Gull, T.~R., Bruhweiler, F., et al.\ 2002, \apj, 581, 1154

\bibitem[Weigelt \& Ebersberger(1986)]{WeigeltEbersberger1986} Weigelt, G., \& Ebersberger, J.\ 1986, \aap, 163, L5

\end{thebibliography}
\end{document}